\documentclass[useAMS,usenatbib]{mn2e}

\usepackage{graphicx}
\usepackage{epsfig}
\usepackage{amssymb,amsmath}
\usepackage{multirow}
\usepackage{mathrsfs}
\usepackage{times}
\usepackage{setspace}
\usepackage{color}

\usepackage[colorlinks, breaklinks, citecolor=blue]{hyperref}

\newcommand{\noopsort}[1]{}

\newcommand{\Q}{\left\langle Q\right\rangle}
\newcommand{\ssb}{\sigma_{\rm SB}}

\usepackage{color}
\definecolor{darkgreen}{rgb}{0.13, 0.55, 0.13}

\title[Simulation of Radiative Feedback]{Resolution Requirements and Resolution Problems in Simulations of Radiative Feedback in Dusty Gas}

\author[Krumholz]{Mark R. Krumholz$^{1,2}$\thanks{mark.krumholz@anu.edu.au}
\\
$^{1}$Research School of Astronomy and Astrophysics, Australian National University, Canberra 2611, ACT, Australia\\
$^{2}$Centre of Excellence for Astronomy in Three Dimensions (ASTRO-3D), Australia
}

\begin{document}
\maketitle
\label{firstpage}
\begin{abstract}
In recent years a number of authors have introduced methods to model the effects of radiation pressure feedback on flows of interstellar and intergalactic gas, and have posited that the forces exerted by stars' radiation output represents an important feedback mechanism capable of halting accretion and thereby regulating star formation. However, numerical simulations have reached widely varying conclusions about the effectiveness of this feedback. In this paper I show that much of the divergence in the literature is a result of failure to obey an important resolution criterion: whether radiation feedback is able to reverse an accretion flow is determined on scales comparable to the dust destruction radius, which is $\lesssim 1000$ AU even for the most luminous stellar sources. Simulations that fail to resolve this scale can produce unphysical results, in many cases leading to a dramatic overestimate of the effectiveness of radiation feedback. Most published simulations of radiation feedback on molecular cloud and galactic scales fail to satisfy this condition. I show how the problem can be circumvented by introducing a new subgrid model that explicitly accounts for momentum balance on unresolved scales, making it possible to simulate dusty accretion flows safely even at low resolution.
\end{abstract}

\begin{keywords}
accretion, accretion disks --- hydrodynamics --- methods: numerical --- radiation: dynamics --- radiative transfer
\end{keywords}


\section{Introduction}
\label{sec:intro}

Newborn massive stars produce intense radiation fields that efficiently heat the interstellar gas and dust around them. While this heating is critical to the observable properties of star-forming regions and the galaxies in which they are embedded, a number of authors have also considered the possibility that the forces exerted by starlight might be significant for gas flows as well. On the scales of individual stars, \citet{larson71a} and \citet{kahn74a} were the first to point out that a sufficiently massive star might exert enough radiation pressure on the gas and dust around it to halt continuing accretion, thereby putting an upper limit on the masses of stars. On the larger scales of star clusters, molecular clouds, and galaxies, \citet{odell67a} and \citet{scoville01a} similarly suggested that radiation pressure might disrupt clouds and ultimately limit the masses of the star clusters to which they give birth. Numerous analytic and semi-analytic models of this phenomenon have been published, considering scales from stellar \citep[e.g.,][]{wolfire86a, wolfire87a, nakano89a, jijina96a} to cluster \citep[e.g.,][]{krumholz09d, fall10a, murray10a, thompson16a, reissl18a}, to galactic \citep[e.g.,][]{murray05a, thompson05a, murray11a, zhang12a, crocker18a}.

However, the interaction between radiation fields and gas is sufficiently complex, and the predictions of analytic models sufficiently uncertain, that investigators in the last two decades have invested significant efforts in numerical study as well. There are a wide variety of numerical methods currently in use, including simple subgrid prescriptions that do not solve the equation of radiative transfer at all \citep{hopkins11a, hopkins18b}, characteristic and hybrid-characteristic methods \citep{kuiper10a, rosen17a}, Monte Carlo methods \citep[e.g.,][]{tsang15a}, and moment methods using the diffusion \citep[e.g.,][]{krumholz07b}, M1 \citep[e.g.,][]{skinner13a, rosdahl15b, kannan18a}, and variable Eddington tensor \citep{davis12a, jiang12a} closures.  A primary goal of this numerical work has been to determine under what circumstances radiation feedback is able to halt accretion onto forming stars -- either individual stars or stellar populations -- and thereby limit the rate and efficiency of star formation. This work has largely been carried out on two parallel tracks, one focusing of the formation of individual stars or small multiple systems, or at most individual star clusters, and a second focusing on the scales of molecular clouds and galaxies. 

On the scales of individual stars, \citet{yorke02a} carried out pioneering 2D radiation-hydrodynamic (RHD) simulations of accretion flows inhibited by radiation pressure, and this was followed by the first 3D RHD simulations by \citet{krumholz09c}. Since then numerous other authors have published RHD simulations that reach stellar masses where radiation pressure begins to have significant impacts on the accretion flow (e.g., \citealt{kuiper10b, kuiper11a, cunningham11a, myers13a, kuiper15a, kuiper16a, klassen16a, rosen16a}; see \citealt{tan14a} for a review). The consensus finding of these studies is that radiation feedback is not particularly effective at halting accretion or limiting stars' ultimate masses. For example, in 2D with laminar initial conditions, \citet{kuiper10b} find star formation efficiencies of 30-50\% (i.e., this fraction of the initial gas mass is ultimately accreted). In 3D using turbulent initial conditions, \citet{rosen16a} set a lower limit of 40\% on the star formation efficiency, with strong hints that this number would rise above 50\% if the simulation continued. No simulations of individual massive star formation published in the past decade have reported star formation efficiencies below $\sim 30\%$.

At larger scales the results have been much less consistent. \citet{hopkins11a, hopkins12b} and \citet{hopkins12f} find that radiation pressure feedback is critical to the regulation of star formation in massive, rapidly star-forming galaxies but is unimportant in more modestly star-forming galaxies, while \citet{ceverino14a} reach exactly the opposite conclusion, that radiation pressure is an effective feedback in low-mass galaxies at modest star formation rates, but not in dense starbursts; \citet{sales14a} and \citet{rosdahl15a} find that the stellar radiation pressure is generally ineffective as a feedback mechanism, at least compared to photoionisation, while \citet{agertz13a} find that it is critical to regulating star formation. Zooming in to individual molecular clouds and clusters, but still working at scales much larger than the simulations of individual stars, \citet{skinner15a}, \citet{raskutti16a, raskutti17a}, \citet{kim17b}, \citet{tsang18a}, and \citet{kim18a}  all find that radiation pressure feedback (as distinct from the effects of photoionisation) is not able to limit star formation efficiencies to less than $30-50\%$ in clouds with column densities typical of observed giant molecular clouds ($\gtrsim 100$ $M_\odot$ pc$^{-2}$), while \citet{grudic18a} find star formation efficiencies an order of magnitude smaller for similar initial conditions; \citet{hopkins18a} attribute this difference to the numerical method used to couple the radiation momentum to the gas.

The divergence between the findings of simulations of the formation of individual massive stars, which uniformly show that accretion is not stopped by radiation pressure and that star formation efficiencies are high, and simulations focusing on the formation of star clusters and galaxies, with their much wider array of outcomes, is at first puzzling. The light to mass ratio of a zero-age stellar population that fully samples the IMF is $\approx 1100$ $L_\odot / M_\odot$ \citep[e.g.,][chapter 7]{krumholz17b}, while that of an individual 60 $M_\odot$ star is $\approx 8300$ $L_\odot / M_\odot$ \citep[e.g.,][]{ekstrom12a}, a factor of $\approx 7$ larger. How can we then make sense of the seemingly paradoxical result that simulations of the formation of individual massive stars nonetheless consistently indicate that stellar radiation feedback is much less effective than do at least some simulations of the formation of star clusters or galactic-scale stellar populations?

Some of the difference in outcome is doubtless due to differences in the choice of initial condition, since the simulations of individual massive star formation generally begin from smaller, denser scales. However, this cannot be the entire explanation. Simple analytic estimates suggest that the effectiveness of radiation feedback should depend on the ratio of surface density to light to mass ratio \citep[e.g.,][]{fall10a}, and many of the massive star formation simulations that reach efficiencies of $\gtrsim 50\%$ with accretion still ongoing start with surface densities that are only a factor of a few larger than cluster simulations where all mass is expelled at much smaller star formation efficiencies. Just to pick one example, \citet{rosen16a}'s run ``TurbRT+FLD" starts at surface density $\Sigma = 1$ g cm$^{-2}$ and forms a single 60 $M_\odot$ star that is still strongly accreting at an efficiency above $40\%$, while \citet{grudic18a}'s ``standard" cluster simulation, with $\Sigma = 1270$ $M_\odot\,\mbox{pc}^{-2}\approx 0.3$ g cm$^{-2}$ and a light to mass ratio 7 times smaller, converts only $30\%$ of its mass to a stars before the cluster's radiation ejects all the remaining mass and causes star formation to cease.

In this paper I show that the key issue is one of resolution: the simulations of individual massive star formation (mostly) satisfy an important resolution criterion, while the larger-scale simulations do not. When the resolution criterion is not satisfied, the results depend sensitively on the exact details of the numerical implementation, and for some implementations the effectiveness of radiation feedback will be drastically overestimated. In \autoref{sec:structure} I begin this demonstration by providing a simplified model for the structure of a radiatively-inhibited dusty accretion flow, which can be solved analytically, and which will provide a baseline against which to test simulations. In \autoref{sec:simulations} I carry out simulations that attempt to reproduce this analytic solution, and show that simulations that fail to satisfy a critical resolution requirement fail to do so. In \autoref{sec:subgrid_model} I describe a subgrid model for radiation feedback that avoids these problems, and successfully reproduces the analytic results even at low resolution. I summarise my findings in \autoref{sec:conclusion}.

\section{Analytic model for dusty, radiation-mediated accretion flows}
\label{sec:structure}

The basic structure of dusty, radiation-inhibited, spherically-symmetric accretion flows onto point sources was first computed a series of seminal papers by \citet{larson71a}, \citet{kahn74a}, \citet{leung75a, leung76a}, and \citet{wolfire86a, wolfire87a}; for a modern update of these papers see \citet{reissl18a}. These papers for the most part involve numerical calculations of the transfer of the stellar radiation field through the dusty envelope, but our goal in this section is to arrive at a simplified model for these flows that is roughly consistent with the numerical results, but is amenable to analytic solution and therefore suitable to serve as the basis for testing numerical methods. The general goal of this analysis is to understand under what circumstances radiation is and is not able to stop accretion flows and drive mass and momentum into the larger environment. For now I omit any discussion of the effects of photoionisation, and I justify this omission below.

\subsection{Opacity and temperature structure}

\begin{figure}
\includegraphics[width=\columnwidth]{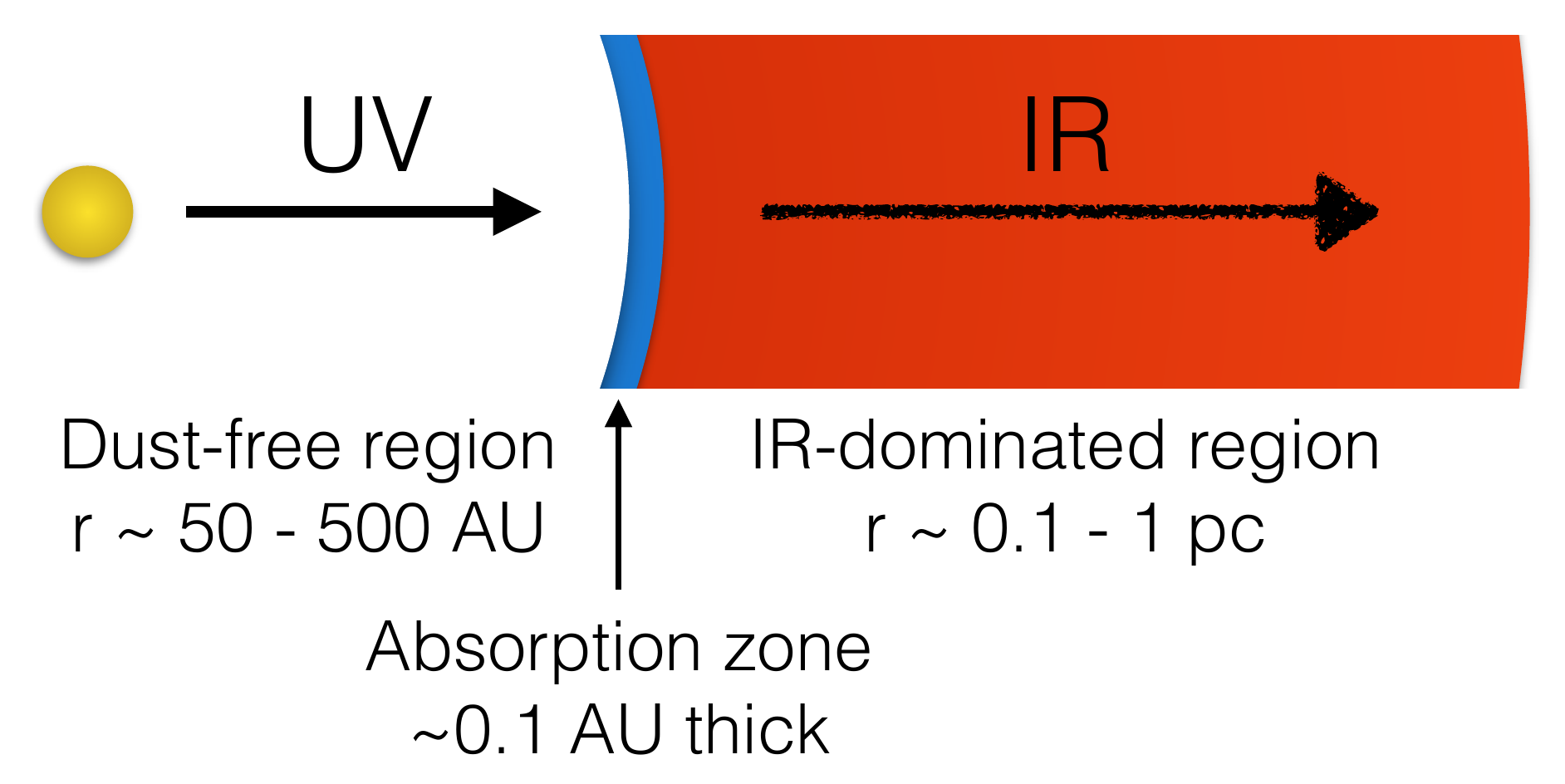}
\caption{
\label{fig:diagram}
Schematic diagram of the temperature and opacity structure of a dusty accretion flow. A central source (yellow circle) creates a dust-free region for tens to hundreds of AU around itself, depending on its luminosity. Ultraviolet stellar photons free-stream through this region, before eventually being absorbed in a very thin shell of dust. In this shell the photons are down-converted to IR, and then they diffuse outward through the dust envelope, before finally diffusing far enough in either radius or frequency to escape.
}
\end{figure}

The temperature and opacity structure around a point source can be roughly divided into three zones, as illustrated in \autoref{fig:diagram}. The first of these is very close to the point source, where the radiation field is intense enough that solid dust grains cannot survive because in thermal equilibrium their temperature would be above the sublimation temperature of their constituent materials. In the absence of dust the only sources of opacity are, depending on the chemical state of the gas, either Thomson scattering by free electrons, absorption of ionising photons by neutral hydrogen, or resonant absorption of photons by molecules \citep[e.g.,][]{malygin14a}. At interstellar densities the flux-mean opacities to starlight provided by these sources are relatively small, $\kappa_F \lesssim 1$ cm$^2$ g$^{-1}$, and thus the region where they dominate is generally optically thin.

As one moves away from the radiation source, the radiation field becomes less intense due to geometric dilution, and at some critical radius dust grains are able to survive. Because the stellar spectrum carries most of its power at wavelengths smaller than the typical grain size, the interaction between the starlight and the grains is close to the limit of geometric optics, and the resulting opacity is large; typical values are $\kappa_F \sim 10^3$ cm$^2$ g$^{-1}$, depending on the stellar spectrum and the grain size distribution \citep{wolfire86a}. The corresponding distance $r_s$ at which grains of radius $a$ and sublimation temperature $T_s$ can survive around a source of luminosity $L$ is given implicitly by the condition of energy balance between absorption and emission at temperature $T_s$:
\begin{equation}
\frac{L}{4\pi r_s^2} \pi a^2 = 4 \pi a^2 \ssb T_s^4 \Q,
\end{equation}
where $\sigma_{\rm SB}$ is the Stefan-Boltzmann constant and $\Q$ is the grain absorption efficiency averaged over a Planck function at temperature $T_s$. Thus the dust sublimation radius is
\begin{equation}
r_s = \sqrt{\frac{L}{16\pi \Q \ssb T_s^4}}
= 780 L_6^{1/2} Q_{-2}^{-1/2} T_{s,3}^{-2}\mbox{ AU},
\end{equation}
where $L_6 = L/10^6$ $L_\odot$, $Q_{-2} = \Q/0.01$, and $T_{s,3} = T_s/1000$ K; typical values for interstellar grains are $Q_{-2}\approx 1$, $T_{s,3}\approx 1.5$. The high opacity of grains to starlight photons guarantees that almost all of the stellar photons are absorbed within a shell of width $\ell \sim (\kappa_F \rho)^{-1} \sim 3 \times 10^{-3}\kappa_{F,3}^{-1} n_{10}^{-1}$ AU, where $\kappa_{F,3} = \kappa_F/10^3$ cm$^2$ g$^{-1}$ and $n_{10}$ is the gas number density in units of $10^{10}$ cm$^{-3}$. This thin absorption region, which has $\ell \ll r_s$, is the second zone.

After the photons are absorbed, they are re-emitted in the infrared. Because the grains are much smaller than the characteristic wavelength for blackbody emission at temperature $T_s$, the flux-mean opacity for the re-emitted photons is much smaller, $\kappa_F \lesssim 10$ cm$^2$ g$^{-1}$. Thus while the region within which the stellar photons are absorbed is of optical depth $\tau_* \sim 1$ to those photons, it is completely transparent, $\tau_{\rm IR} \sim 0.01$, to the re-emitted IR photons. However, because there is generally a large column of material outside the absorption region, the IR photons generally do not immediately escape to infinity. Instead, they escape the absorption region but then must diffuse outward through the remainder of the dusty accretion flow, experiencing repeated absorptions and re-emissions that shift them to ever-lower frequencies and result in lower flux-mean opacities, until they finally escape. The flux-mean opacity in this diffusion region is a complex function of temperature, governed by temperature-dependent sublimation and condensation of different grain species, but it can be roughly approximated as \citep{semenov03a}
\begin{equation}
\label{eq:IR_opacity}
\kappa_{\rm IR} \approx \kappa_{\rm IR,0}
\left\{
\begin{array}{ll}
(T/T_0)^2, & T < T_0 \\
1, & T_0 \leq T < T_s \\
0, & T_s \leq T
\end{array}
\right.
\end{equation}
where $T$ is the radiation temperature, $\kappa_{\rm IR,0} \approx 7$ cm$^2$ g$^{-1}$, $T_0 \approx 150$ K. The radiation temperature is similarly a complex function of opacity, which for full accuracy must be obtained numerically. However, it can reasonably be approximated as a powerlaw in radius \citep[e.g.,][]{wolfire86a, chakrabarti05a, chakrabarti08a},
\begin{equation}
\label{eq:tprof}
T \approx \phi T_s \left(\frac{r}{r_s}\right)^{-k_T}
\end{equation}
where $k_T \approx 0.5$ and $\phi \approx 0.3$.

\subsection{Kinematic structure}

Next let us consider the kinematic structure of the flow, which is determined by the balance between gravitational and radiative forces; since dusty accretion flows near point sources are generally highly supersonic, we can neglect pressure forces. The gravitational force per unit mass is simply $G (M_* + M_r)/r^2$, where $M_*$ is the mass of the central source and $M_r$ is the gas mass interior to radius $r$. For the purpose of calculating the radiation force, I assume that the dust temperature obeys \autoref{eq:tprof}. The luminosity $L$ passing through any given radius is constant, and can be divided up into a direct starlight component of luminosity $L_*$ and a dust-processed infrared component of luminosity $L_{\rm IR} = L - L_*$; the opacities of the material to these two components are
\begin{equation}
\kappa_* =
\left\{
\begin{array}{ll}
\kappa_{*,0}, & T < T_s \\
0, & T \geq T_s
\end{array}
\right..
\end{equation}
and $\kappa_{\rm IR}$ (\autoref{eq:IR_opacity}), respectively. Combining these considerations, we can write the full equation of motion for a fluid element at radius $r$ as
\begin{equation}
\frac{dv}{dt} = -\frac{G(M_*+M_r)}{r^2} + \frac{L}{4\pi r^2 c}  \left[\kappa_* e^{-\tau_*} + \kappa_{\rm IR} \left(1 - e^{-\tau_*}\right)\right]
\label{eq:eom_dim}
\end{equation}
where $dv/dt$ is the Lagrangian derivative of the velocity,
\begin{eqnarray}
M_r & = & \int_0^r 4\pi r'^2 \rho \, dr' \\
\tau_* & = & \int_{r_s}^r \kappa_* \rho \, dr'
\end{eqnarray}
are the mass interior to radius $r$ and the optical depth to starlight photons at radius $r$ respectively, and $\rho$ is the gas density. Note that $\kappa_{\rm IR}$ and $\kappa_*$ are both functions of temperature and thus of position. In \autoref{eq:eom_dim}, the first term inside the square brackets represents the force exerted by the direct starlight field, carrying a luminosity $L_* = Le^{-\tau_*}$, while the second represents the force exerted by the reprocessed infrared radiation field, carrying a luminosity $L_{\rm IR} = L(1-e^{-\tau_*})$. 

It is convenient to non-dimensionalise this equation via the change of variables
\begin{equation}
x = \frac{r}{r_s} \quad u = \frac{v}{\sqrt{G M_* / r_s}}
\quad s = \frac{t}{\sqrt{r_s^3/G M_*}}
\quad b = \frac{\rho}{M_*/r_s^3},
\end{equation}
which produces
\begin{eqnarray}
\frac{du}{ds} & = & -\frac{1+m_x}{x^2} + \frac{f_E}{x^2} \left[k_* e^{-\tau_*} + k_{\rm IR}\left(1-e^{-\tau_*}\right)\right]
\label{eq:eom}
\\
m_x & = & \int_0^x 4\pi x'^2 b \, dx' \\
\tau_* & = & f_\tau \int_1^{\max(1,x)} b \, dx' \\
k_* & = & 
\left\{
\begin{array}{ll}
0, & x < 1 \\
\eta, & x \geq 1
\end{array}
\right.
\\
k_{\rm IR} & = &
\left\{
\begin{array}{ll}
0, & x < 1 \\
1, & 1 \leq x < x_T \\
(x/x_T)^{-2k_T}, & x \geq x_T
\end{array}
\right..
\end{eqnarray}
The dimensionless ratios appearing in this equation are
\begin{eqnarray}
\eta & = & \frac{\kappa_{*,0}}{\kappa_{\rm IR,0}} \approx 100 \\
f_E & = & \frac{\kappa_{\rm IR,0} L}{4\pi G M_* c} = 0.78 \kappa_1 \Psi_3 \\
f_\tau & = & \frac{16\pi \langle Q\rangle \sigma_{\rm SB} T_s^4 \eta \kappa_{\rm IR,0}}{\Psi}
= 1.4\times 10^7 \frac{\kappa_1 Q_{-2} T_{s,3}^4}{\Psi_3} \\
x_T & = & \left(\phi\frac{T_s}{T_0}\right)^{1/k_T} \approx 10,
\end{eqnarray}
and physically they represent the ratio of dust opacities for stellar photons and IR photons, the Eddington ratio for the maximum IR opacity, the dimensionless optical depth per unit column to stellar photons, and the ratio of the radius where the temperature drops to $T_0$ to the dust sublimation radius, respectively. In the numerical evaluations, $\kappa_1 = \kappa_{\rm IR,0}/10$ cm$^2$ g$^{-1}$ and $\Psi = L/M_* = 10^3 (L_\odot/M_\odot) \Psi_3$, so $\Psi_3 \approx 1$ for a zero-age stellar population that samples the IMF, and $\Psi_3 \approx 10$ for the most massive individual stars.

Since our goal is to determine under what conditions this equation of motion permits a steady inflow solution with mass moving inward at a time- and space-independent rate $\dot{M} = 4\pi r^2 \rho |v|$, and what should happen when no such solution is possible, it is helpful to introduce an additional dimensionless parameter. I define
\begin{equation}
f_p = \frac{\dot{M} (GM_*/r_s)^{1/2}}{L/c} = 1.6 \dot{M}_{-3} Q_{-2}^{1/4} T_{s,3} L_6^{-3/4} \Psi_3^{-1/2}.
\end{equation}
as the ratio of the momentum carried by a flow of gas freely-falling in stellar gravity to radius $r_s$ to the momentum carried by the stellar radiation field; here $\dot{M}_{-3} = \dot{M}/10^{-3}$ $M_\odot$ yr$^{-1}$. With this definition, one can equivalently express the dimensionless density, mass, and optical depth for any steady-state inflow solution as
\begin{eqnarray}
\label{eq:density}
b & = & \frac{\eta f_E f_p}{f_\tau} \frac{1}{|u| x^2} \\
\label{eq:mx}
m_x & = & 4\pi \frac{f_E f_p \eta}{f_\tau} \int_0^x \frac{1}{|u|} dx' \\
\tau_* & = & \eta f_E f_p \int_1^{\max(1,x)} \frac{1}{|u| x'^2} dx'.
\label{eq:tau_star}
\end{eqnarray}

\subsection{Inflow solutions}
\label{ssec:inflow_sol}

We are now in a position to determine when a steady inflow solution is possible. First note that, since $f_E f_p \eta/f_\tau \ll 1$ for our fiducial parameters (or indeed for any plausible set of physical parameters, since $f_\tau$ is so large), we expect the gas mass at $x\sim 1$ to be negligible in comparison to the stellar mass, and we can therefore drop the term $m_x$ in \autoref{eq:eom} for the purposes of this analysis; I confirm this explicitly below.

Consider the region inside the dust destruction front, $x<1$, where the equation of motion is simply
\begin{equation}
\frac{du}{ds} = u\frac{du}{dx} = -\frac{1}{x^2}.
\end{equation}
If the velocity is $u_{1^-}$ just inside the dust destruction front at $x=1$, then the we can solve the equation of motion immediately to find
\begin{equation}
\label{eq:u_UV}
u = - \sqrt{u_{1^-}^2 + 2\left(\frac{1}{x}-1\right)}.
\end{equation}

Next consider the region near $x=1$, where UV photons are absorbed. We can solve for the flow in this region by making two important observations. First, since $\eta f_E f_p \gg 1$, the coefficient in front of the integral that defines $\tau_*$ (\autoref{eq:tau_star}) is very large, $\sim 100$. Thus we will have $\tau_* \gg 1$ for any $x$ even slightly larger than unity, implying that the transition from $\tau_* = 0$ to $\tau_* \gg 1$ occurs entirely within a thin region near $x=1$, consistent with the sketch in \autoref{fig:diagram}. All UV photons will be absorbed in this thin region. Second, since $\eta \approx 100$, within the region where $\tau_* \lesssim 1$, the middle term on the right hand side of \autoref{eq:eom} that represents the UV radiation force, $f_E k_* e^{-\tau_*}/x^2$, is roughly two orders of magnitude larger than either the gravity term (the first term) or the IR force term (the third term). The UV force term does not become comparable to the others until $\tau_* \gtrsim 5$. Consequently, in the thin region where $\tau_*$ is transitioning from $0$ to $\gtrsim 1$, we can to good approximation drop the gravity and IR force terms in the equation of motion, obtaining 
\begin{equation}
u \frac{du}{dx} = \frac{\eta f_E}{x^2} \exp\left(-\eta f_p f_E \int_1^x \frac{1}{|u| x'^2} dx'\right).
\end{equation}
We can integrate this by making a change of variables from $x$ to $\tau_*$, which yields
\begin{equation}
\frac{du}{d\tau_*} = -\frac{1}{f_p} e^{-\tau_*}.
\end{equation}
Integrating from $\tau_* = 0$ to $\tau_* \gg 1$, we find that the velocity $u_{1^-}$ just inward of the absorption region is related to the velocity $u_{1^+}$ just outside it by
\begin{equation}
u_{1^-} = u_{1^+} + \frac{1}{f_p}.
\end{equation}
This result becomes exact as $\eta \rightarrow \infty$, and the UV absorption region becomes arbitrarily thin. Thus we conclude that an inflow, with $u_{1^-} \leq 0$, can exist only if $u_{1^+} < -1/f_p$.

Finally consider the region where $\tau_* \gg 1$, within which the equation of motion becomes
\begin{equation}
u \frac{du}{dx} = \frac{1}{x^2} \left\{-1 + f_E \min\left[1,\left(x/x_T\right)^{-2 k_T}\right]\right\}
\end{equation}
We can then solve for $u$ directly, subject to the boundary condition that $u\rightarrow 0$ as $x\rightarrow \infty$. The result is
\begin{equation}
\label{eq:u_IR}
u = -\sqrt{\frac{2}{x}}
\left\{
\begin{array}{ll}
\sqrt{1 - \frac{f_E}{1+2 k_T}\left(1+2k_T -2 k_T \frac{x}{x_T}\right)}, & 1 < x \leq x_T \\
\sqrt{1 - \frac{f_E}{1+2 k_T}\left(\frac{x}{x_T}\right)^{-2k_T}}, & x > x_T
\end{array}
\right..
\end{equation}
Thus our condition that $u_{1^+} < -1/f_p$ is satisfied only if
\begin{eqnarray}
f_p > f_{p,\rm crit} & \equiv & \left\{2\left[1-\left(1 - \frac{2 k_T}{x_T+2 x_T k_T}\right)f_E\right]\right\}^{-1/2}
\nonumber \\
& \approx & \left[2\left(1-0.95 f_E\right)\right]^{-1/2}
\label{eq:fp_con}
\end{eqnarray}
where the numerical evaluation is for $k_T = 0.5$ and $x_T = 10$. In dimensional terms, for a zero age stellar population and our fiducial opacity choice ($\Psi_3 = 1.1$, $\kappa_1 = 0.7$), we have $f_E = 0.6$, and thus we can express the condition for accretion to be possible as $f_p > 1.1$, or, in dimensional terms,
\begin{equation}
\dot{M} > 6.5\times 10^{-4} Q_{-2}^{-1/4} T_{s,3}^{-1} L_6^{3/4}\,M_\odot\mbox{ yr}^{-1}.
\end{equation}

We can also use this solution to verify directly that $m_x$ is indeed negligible. The maximum possible mass of gas at a given accretion rate corresponds to the minimum possible gas velocity. This is achieved when $f_E$ and $f_p$ are such that the inflow condition \autoref{eq:fp_con} is just satisfied, and $u_{1^-} = 0$. In this case we can evaluate the integral in \autoref{eq:mx} for $m_x$, using \autoref{eq:u_UV} inside $x=1$ and \autoref{eq:u_IR} outside $x=1$. Doing so we find that $m_x \lesssim 1$ at $x=100$ for all $f_p \lesssim 100$, and that at $f_p = 10$ (close to the upper limit we expect for realistic parameters) $m_x \approx 0.1$ at $x=100$. Thus for realistic values of $f_p$, the radius at which gas self-gravity becomes significant is much larger than the radii at which the great majority of the radiative and gravitational acceleration occurs, and where the balance between the two is determined. We are therefore justified in ignoring self-gravity for the purposes of determining when accretion is possible.

What happens if no steady inflow is possible, $f_p < f_{p,\rm crit}$? In this case the velocity must become 0 or positive at $r=r_s$, leading to formation of a shock. Since gas will cool rapidly behind the shock, there will be a large density jump, and as a result mass must accumulate in a dense shell. Since the net force on the shell will be outward, the shell will begin to move out and sweep up the gas around it. The flow in that case will approach the analytic similarity solution for radiation-driven spherical shells derived by \citet{krumholz09d}. It will deviate from this solution only once the surface density through the shell becomes large enough that the gravitational force on the shell exceeds the radiative force:
\begin{eqnarray}
\Sigma_{\rm sh} & = & \frac{L}{4 \pi G (M_* + M_{\rm sh}/2) c} 
\nonumber \\
& = &
0.077\, \Psi_3 \left(1+\frac{M_{\rm sh}}{2M_*}\right)^{-1}\mbox{ g cm}^{-2},
\end{eqnarray}
where $\Sigma_{\rm sh}$ is the shell mass per unit area, and $M_{\rm sh}$ is the shell mass.

\subsection{Photoionisation}

Thus far in this calculation I have neglected pressure forces under the assumption that accretion flows are highly supersonic. However, this assumption might break down if gas becomes ionised by photons from the central source, which would raise its temperature to $\approx 10^4$ K, and its sound speed to $\approx 10$ km s$^{-1}$. I now show that this does not generally happen, because if $f_p$ is large enough to admit an inflow solution, then it is also large enough that we can safely neglect the pressure of photoionised gas; the calculation here closely follows that of \citet{walmsley95a}, and echoes the conclusions previously drawn by \citet{keto02a, keto03a}.

First note that inside the dust sublimation radius the magnitude of the velocity is bounded above by $|v| = v_{\rm ff} = \sqrt{GM_*/r}$ (i.e., the velocity cannot exceed the free-fall speed from infinity), and thus the density obeys $\rho > \dot{M}/(4\pi r^2 v_{\rm ff})$. Photoionisation balance requires that, if the central source has an ionising luminosity $S$ (measured in photons per unit time), and the photoionised region has an inner radius $r_0$, it must have an outer radius $r_i$ given implicitly by the condition
\begin{equation}
S = \int_{r_0}^{r_i} 4\pi r^2 \alpha_{\rm B} x_e \left(\frac{\rho}{\mu_{\rm H} m_{\rm H}}\right)^2 \, dr,
\end{equation}
where $\alpha_B \approx 2.54 \times 10^{-13} (T/10^4\mbox{ K})^{-0.82}$ cm$^3$ s$^{-1}$ is the case B recombination coefficient \citep{draine11a}, $x_e$ is the free electron abundance per H nucleus ($\approx 1.1$ if He is singly-ionised, $\approx 1.2$ if it is doubly-ionised) and $\mu_{\rm H}$ is the mean mass per H nucleus in units of the hydrogen mass $m_{\rm H}$; for standard cosmic abundance, $\mu_{\rm H} = 1.4$. Inserting our lower limit on $\rho$ yields an upper limit on $r_i$:
\begin{eqnarray}
r_i & < & r_0 \exp\left(\frac{8\pi \mu_{\rm H}^2 m_{\rm H}^2 G M_* S}{\alpha_B x_e \dot{M}}\right) \\
& = & r_0 \exp\left[\frac{2 \pi^{3/2} \left(4\mu_{\rm H} m_{\rm H} c G T_s\right)^2 \sqrt{Q L \sigma_{\rm SB}}}{f_p^2 x_e \alpha_B \gamma \Psi^2}\right] \\
& = & r_0 \exp\left(1.8 L_6^{1/2} Q_{-2}^{1/2} T_{s,3}^2 \Psi_3^{-2} f_p^{-2} \right)
\label{eq:ri}
\end{eqnarray}
Here $\gamma = L/S$ is the mean energy radiated by the source per ionising photon emitted ($\gamma \approx 3.2$ Ryd for a standard IMF -- \citealt{fall10a}), and the numerical evaluation in the final line uses $T = 10^4$ K and $x_e = 1.1$ for the photoionised gas.

The fact that the quantity in parentheses in \autoref{eq:ri} is of order unity implies that, if $f_p$ is large enough to permit accretion ($f_p \gtrsim 1$), the photoionised region will be confined to a relatively small radial extent. For our fiducial parameters, its ratio of inner to outer radius, $r_i/r_0$, will not exceed a factor of a few. Thus unless some other mechanism (e.g., a stellar wind bubble) pushes the inner edge of a photoionised region out to within a factor of a few of $r_s$, the photoionised region will be confined entirely to well within $r_s$, simply because the density in the accretion flow is high.

In turn, however, this implies stringent limits on the dynamical importance of the photoionised gas. The physical inflow velocity is
\begin{equation}
v = \sqrt{\frac{GM_*}{r_s}} u = 29 u T_{s,3} Q_{-2}^{1/4} L_6^{1/4} \Psi_3^{-1/3} \mbox{ km s}^{-1}.
\end{equation}
Thus for the inflow velocity to be subsonic with respect to the sound speed in photoionised gas, we must have $|u| \lesssim 0.3$ in a region where the gas is photoionised. However, inside $r_s$ our inflow solution (\autoref{eq:u_UV}) implies $|u| \geq \sqrt{2(x^{-1}-1)}$, with equality holding if $u_{1^-} = 0$, i.e., if the radiation force is able to stop the flow completely at the dust sublimation front. This in turn means that $|u| > 0.3$ at all radii $x \lesssim 0.95$. In words, even if we consider a flow that only barely carries enough momentum to accrete, and thus comes to nearly a dead stop at the dust sublimation front, the flow will accelerate to be faster than the ionised gas sound speed after moving inward in radius only 5\% further. However, we have just seen that the ionisation front will generally be located much farther inward than this, unless there is some mechanism other that photoionisation alone to push it outward. Thus by the time the accretion flow falls to radii small enough to be ionised, it will also be moving much too fast for the resulting increase in pressure to alter its trajectory much. For this reason, ionisation will not significantly change the condition for inflow to occur.\footnote{This does not mean that ionised gas pressure is not important. If the accretion rate drops low enough for the accretion flow to be reversed, then the pressure of the expanding photoionised bubble may be very important for the subsequent dynamics. This will depend on the ratio of ionised gas to photon pressure, as discussed for example by \citet{krumholz09d} and \citet{draine11b}. The point here is simply that consideration of ionised gas pressure does not alter the conditions that determine inflow versus ejection.}

\subsection{Summary of analytic results and their limitations}
\label{ssec:analytic_summary}

We can now summarise the key findings of our analytic investigation, which we will test the ability of numerical simulations to recover. These are:
\begin{enumerate}
\item For fixed dust properties, whether radiation pressure is sufficient to halt an accretion flow is determined primarily by two dimensionless parameters: $f_E$, the Eddington ratio computed for the maximum dust opacity, and $f_p$ the ratio of inflow momentum to radiation momentum. The former is primarily sensitive to the light to mass ratio of the driving source, and is $\sim 0.5$ for fully sampled stellar population. The latter is primarily sensitive to accretion rate and stellar luminosity.
\item If $f_p\gtrsim 1$ (for $f_E\lesssim 0.6$), the accretion flow carries enough momentum to crush the radiation field, so that almost all the radiation momentum is deposited in a small region close to the source, but is then advected back into the source with the accretion flow. In this case no radiation feedback is felt far from the source. For Milky Way-like dust, this condition prevails if the accretion rate $\dot{M} \gtrsim 5\times 10^{-4} L_6^{3/4}$ $M_\odot$ yr$^{-1}$, where $L_6$ is the source luminosity in units of $10^6$ $L_\odot$.
\end{enumerate}

Finally, it is important to emphasise a major caveat of these results, which is that they are for spherically-symmetric flows. While this simplification is necessary in order to obtain analytic results that can serve as a testbed for simulations, it is worth considering the limitations of this approach. First the case where accretion is still onto a single point source, so the output radiation field is spherically symmetric or nearly so, but the in reality the flow pattern need not be. In such a case, the assumption of spherical symmetry probably matters relatively little in the regime $f_E < 1$. In this case, the balance between inflow and outflow is determined mainly by the direct stellar radiation field, which is spherical. One can then simply apply these results on a sector-by-sector basis. On the other hand, for $f_E > 1$, as can happen for very massive stars, numerous simulations over many years have shown that spherical symmetry is a poor assumption, and that breaking of that symmetry allows inflow to continue even when a spherically-symmetric calculation suggests it should halt \citep[e.g.,][]{krumholz09c, kuiper11a, rosen16a}. The focus of this paper is on the $f_E < 1$ regime.

Now consider the case where there are multiple point sources of radiation rather than a single one. If the point sources were separated by a distance $\lesssim r_s$, then one could reasonably approximate them as a single source within a single dust sublimation front. However, even the most compact and massive known star clusters, e.g.~R136, tend to have separations of $\gg 1000$ AU between O stars \citep[e.g.,][]{massey98b}. Thus we are likely to be in the opposite limit where the interstellar separations are $\gg r_s$. In this case the analysis above should be applied to each point source separately. In particular, if the accretion flow onto each individual point sources is high enough to produce $f_p \gtrsim 1$ for its luminosity, then the momentum of each source will be advected back onto it, and there will be no interaction between the radiative momentum deposition from the different sources.

\section{Discretisation and numerical tests}
\label{sec:simulations}

The full source code for the numerical scheme I describe below, and for all the calculations I perform with it in subsequent sections, is available from \url{https://bitbucket.org/krumholz/dusty_resolution_tests/}.

\subsection{Numerical scheme}
\label{ssec:scheme}

Having established analytically under what conditions an inflow should and should not be possible, I next investigate the ability of simulations with finite resolution to reproduce these results. For the purposes of this test I consider a spherically-symmetric gas whose Lagrangian equation of motion is \autoref{eq:eom} with two extra terms representing pressure and viscous forces:
\begin{eqnarray}
\frac{du}{ds} & = & -\frac{1+m_x}{x^2} + \frac{f_E}{x^2} \left[k_* e^{-\tau_*} + k_{\rm IR}\left(1-e^{-\tau_*}\right)\right] 
\nonumber \\
& & \qquad {} + a^2\frac{db}{dx} + \nu \nabla^2 u.
\label{eq:eom_sim}
\end{eqnarray}
Here $a$ is the dimensionless sound speed, $b$ is the dimensionless density, and $\nu$ is the dimensionless viscosity. These extra terms are negligibly small most places in the flow, because I will choose the coefficients $a$ and $\nu$ to be small, but they become non-negligible in shocks, and they are required to ensure that shocks develop properly in the simulations.\footnote{Note that in taking $a$ to be constant, I am implicitly treating the gas as isothermal, which is not consistent with the assumed temperature profile. However, since I will be taking $a$ to be so small that the pressure term is negligible except within shocks, there is no reason to treat it more accurately. In the simulations I carry out in this paper, it is best to think of the pressure term as an artificial pressure that, in conjunction with an artificial viscosity, makes it possible to resolve shocks.}

I solve this equation using a simple one-dimensional, spherical, Lagrangian scheme, which combines aspects of the methods of \citet{cioffi88a} and \citet{mezzacappa93a}. I consider a series of cell edges with mass coordinate $m_i$, denoting the mass enclosed. The cells are uniformly spaced in mass, with $m_{i+1} - m_i = \Delta m$. Each cell edge has a radial coordinate $x_i$ which moves with velocity $u_i$. The time-derivative of the velocities is
\begin{equation}
\label{eq:eom_discrete}
\frac{du_i}{ds} = f_{{\rm grav},i} + f_{*,i} + f_{{\rm IR},i} + f_{{\rm pres},i} + f_{{\rm visc},i}
\end{equation}
where the five terms on the right hand side represent the force per unit mass from gravity, direct stellar radiation, infrared radiation, gas pressure, and viscosity, respectively. The discretised gravitational, pressure, and viscous force terms are standard for Lagrangian methods:
\begin{eqnarray}
f_{{\rm grav},i} & = & -\frac{1 + i\,\Delta m}{x_i^2} \\
f_{{\rm pres},i} & = & 4\pi x_i^2 a^2 \frac{b_{i+1/2} - b_{i-1/2}}{\Delta m} \\
f_{{\rm visc},i} & = & 4\pi x_i^2 \frac{q_{i+1/2}-q_{i-1/2}}{\Delta m}
\end{eqnarray}
Here $b_{i+1/2} = 3\, \Delta m/4\pi(x_{i+1}^3-x_{i}^3)$ denotes the mean density of the cell that lies between edges $m_i$ and $m_{i+1}$, and
\begin{equation}
q_{i+1/2} = 
\left\{
\begin{array}{ll}
\nu b_{i+1/2} \left(u_{i+1} - u_i\right)^2, & u_{i+1} < u_i \\
0, & u_{i+1} \geq u_i
\end{array}
\right.
\end{equation}
is the viscous momentum flux per unit mass, calculated as a standard \citet{von-neumann50a} quadratic artificial viscosity. I use $a^2 = 10^{-3}$ and $\nu=2$ for all simulations.

I discretise the IR radiation force as
\begin{equation}
f_{{\rm IR},i} = \frac{f_E}{x_i^2} \left(1 - e^{-\tau_{*,i}}\right) \min\left[1, \left(\frac{x_i}{x_T}\right)^{-2k_T}\right]
\end{equation}
where 
\begin{equation}
\tau_*(x_i) = f_\tau \sum_{j=0}^{i-1} b_{j+1/2} \left[\max\left(x_{j+1},1\right) - \max\left(x_j,1\right)\right]
\end{equation}
is the optical depth to stellar photons to position $x_i$. The direct stellar radiation field term requires a bit more care. At any given radius $x$ the force per unit mass is $f_* = \eta f_E e^{-\tau_*}/x^2$, but $f_\tau$ is so large that even in a high resolution calculation it is generally not practical to choose $\Delta m$ small enough so that the length scale over which $\tau_*$ goes from 0 to $\gg 1$ is resolved by more than a few cells. Thus one must explicitly average over cells in order to ensure that the force is calculated correctly and the amount of radial momentum delivered to the flow adds up to exactly $L/c$. The average force per unit mass exerted on the material between cell edges $i$ and $i+1$
\begin{eqnarray}
\langle f_*\rangle_{i+1/2} & = & \frac{3}{4\pi \left(x_{i+1}^3-x_i^3\right)} \int_{x_{i}}^{x_{i+1}} 4\pi x^2 f_* \, dx \\
& = & \frac{4\pi \eta f_E}{f_\tau\, \Delta m} \left(e^{-\tau_{*,i}} - e^{-\tau_{*,i+1}}\right).
\end{eqnarray}
\citet{hopkins18a} point out that, for an Eulerian coordinate system where the momenta carried by stellar photons emitted in different directions can cancel, the choice of where to deposit this momentum can have significant consequences for the outcome. For the spherical grid I use here this problem does not occur, and in any event the fix proposed by \citet{hopkins18a} is not relevant for the resolution problem I discuss below. For the purposes of the simulations I carry out, I deposit the momentum on the inner cell face, which amounts to taking
\begin{equation}
f_{*,i} = \langle f_*\rangle_{i+1/2} = \frac{4\pi \eta f_E}{f_\tau\, \Delta m} \left(e^{-\tau_{*,i}} - e^{-\tau_{*,i+1}}\right).
\end{equation}

I advance the simulation in time using a 2nd-order accurate time stepping scheme, taken from \citet{cioffi88a}. Formally, let
\begin{equation}
\mathbf{Z} =
\left(
\begin{array}{c}
x_i \\
u_i
\end{array}
\right)
\end{equation}
be the state vector for the system, which evolves following
\begin{equation}
\dot{\mathbf{Z}} =
\left(
\begin{array}{c}
u_i \\
(du/ds)_i
\end{array}
\right),
\end{equation}
where the dot indicates differentiation with respect to the dimensionless time $s$, and $(du/ds)_i$ is evaluated from \autoref{eq:eom_discrete}. To advance the dimensionless time through a step $\Delta s$, I carry out the following update cycle, where $\mathbf{Z}^{n}$ denotes the state at dimensionless time $s_n$ and $\mathbf{Z}^{n+1}$ the state at dimensionless time $s_{n+1} = s_n + \Delta s$:
\begin{eqnarray}
\mathbf{Z}^{n+1/2} & = & \mathbf{Z}^{n} + \frac{\Delta s}{2}\dot{\mathbf{Z}}^{n} \\
\mathbf{Z}^{n+1,*} & = & \mathbf{Z}^{n} + \Delta s \, \dot{\mathbf{Z}}^{n+1/2} \\
\mathbf{Z}^{n+1} & = & \mathbf{Z}^{n} + \frac{\Delta s}{2} \left(\dot{\mathbf{Z}}^{n} + \dot{\mathbf{Z}}^{n+1,*}\right). 
\end{eqnarray}
I set the time step to
\begin{equation}
\Delta s = C \min \frac{-\left(\left|\Delta u_i\right| + a\right) + \sqrt{\left(\left|\Delta u_i\right|+a\right)^2 + 2 \left|\Delta \dot{u}_i\right| \Delta x_i}}{\left|\Delta\dot{u}_i\right|},
\end{equation}
where $C$ is the CFL number (set to 0.5 for all calculations here), $\Delta x_i = x_{i+1} - x_i$, $\Delta u_i = u_{i+1}-u_i$, and $\Delta \dot{u}_i = \dot{u}_{i+1} - \dot{u}_i$. Note that this is just the generalisation of the usual Lagrangian Courant condition to include the effects of an acceleration during the time step, and that for $\left|\Delta \dot{u}\right| \Delta x \ll \left|\Delta u\right|$ this condition reduces the usual Courant time step. This generalisation is helpful because the acceleration at the dust destruction front can be extremely large, causing the ordinary CFL condition to be insufficient to maintain stability. Even with this addition, the update on occasion allows two shells to cross, so that $x_i \geq x_{i+1}$ either at the end of a time step or during one of the intermediate updates. If the code detects this condition, it simply reduces the time step size $\Delta s$ and retries the advance until the step succeeds. Finally, note that cells that reach $x < 1$ will fall to the origin in finite time, and this could cause the calculation to grind to a halt because the acceleration diverges as $x\rightarrow 0$. To avoid this I remove from the calculation any cell edge that falls to below $x = 0.5$. I retain the cell's mass for the purposes of calculating the gravitational force, but do not further update its position or velocity.

\subsection{Resolution study}

\subsubsection{High resolution}
\label{sssec:high_res}

To test whether simulations can recover the analytic solution, I consider a case with all parameters set to their fiducial values for a zero-age stellar population and Milky Way dust: $\Psi = 1100$ $L_\odot$ $M_\odot^{-1}$, $T_s = 1500$ K, $T_0 = 150$ K, $\langle Q\rangle = 0.01$, $\kappa_{\rm IR,0} = 7$ cm$^2$ g$^{-1}$, $\kappa_* = 700$ cm$^2$ g$^{-1}$, $\phi = 0.4$, $k_T = 0.5$. For these choices, the dimensionless parameters for the problem are $f_E = 0.59$, $f_\tau = 4.8\times 10^7$, $x_T = 16$, and $\eta = 100$, and the dimensionless momentum inflow rate required to allow accretion is $f_{p,\rm crit} = 1.1$. 

I first consider a simulation that is able to resolve the location of the dust destruction front. For the parameters specified, and a momentum flux $f_p = 2$, the total mass interior to the dust sublimation front for the analytic solution is $m_s = 2.5\times 10^{-5}$. Thus for the fiducial high resolution test I adopt $\Delta m = 10^{-6}$, so the dust sublimation region is resolved by $\approx 25$ zones. I initialise the simulation by placing the innermost zone at $x_0 = 1.0$, initialising the velocities and densities of all subsequent zones to the analytic solution for $x>1$. The procedure is as follows: starting from the first zone edge, I integrate \autoref{eq:density} numerically using the velocity $u$ taken from \autoref{eq:u_IR} until I reach a radius where the enclosed mass is $\Delta m$. This becomes the initial location of the next cell edge, and I use \autoref{eq:u_IR} to initialise its velocity. I use this procedure to initialise a total of 500 cell edges, for the cases $f_p = 0.4, 0.8, 1.2, 1.4$, and $2.0$.\footnote{Note that the analytic solution becomes undefined inside $x = 1$ for $f_p < f_{p,\rm crit}$, but since I only initialise the flow in the region $x>1$ this does not create any problems.} I then simulate each case to time $s=1$; recall that $s=1$ corresponds to roughly the free-fall time from $r_s$ to the central point mass, so the simulations are run for a time long enough to follow material that starts are the dust sublimation front all or most of the way onto the central object. For reference, if the central object luminosity is $L = 10^6$ $L_\odot$, the mass resolution of these simulations is $9.1\times 10^{-4}$ $M_\odot$, the run time is 34 yr, the dust sublimation radius is $r_s = 340$ AU, and the accretion rates range from $1.7\times 10^{-4}$ $M_\odot$ yr$^{-1}$ (for $f_p = 0.4$) to $8.4\times 10^{-4}$ $M_\odot$ yr$^{-1}$ (for $f_p = 2.0$).

\begin{figure}
\includegraphics[width=\columnwidth]{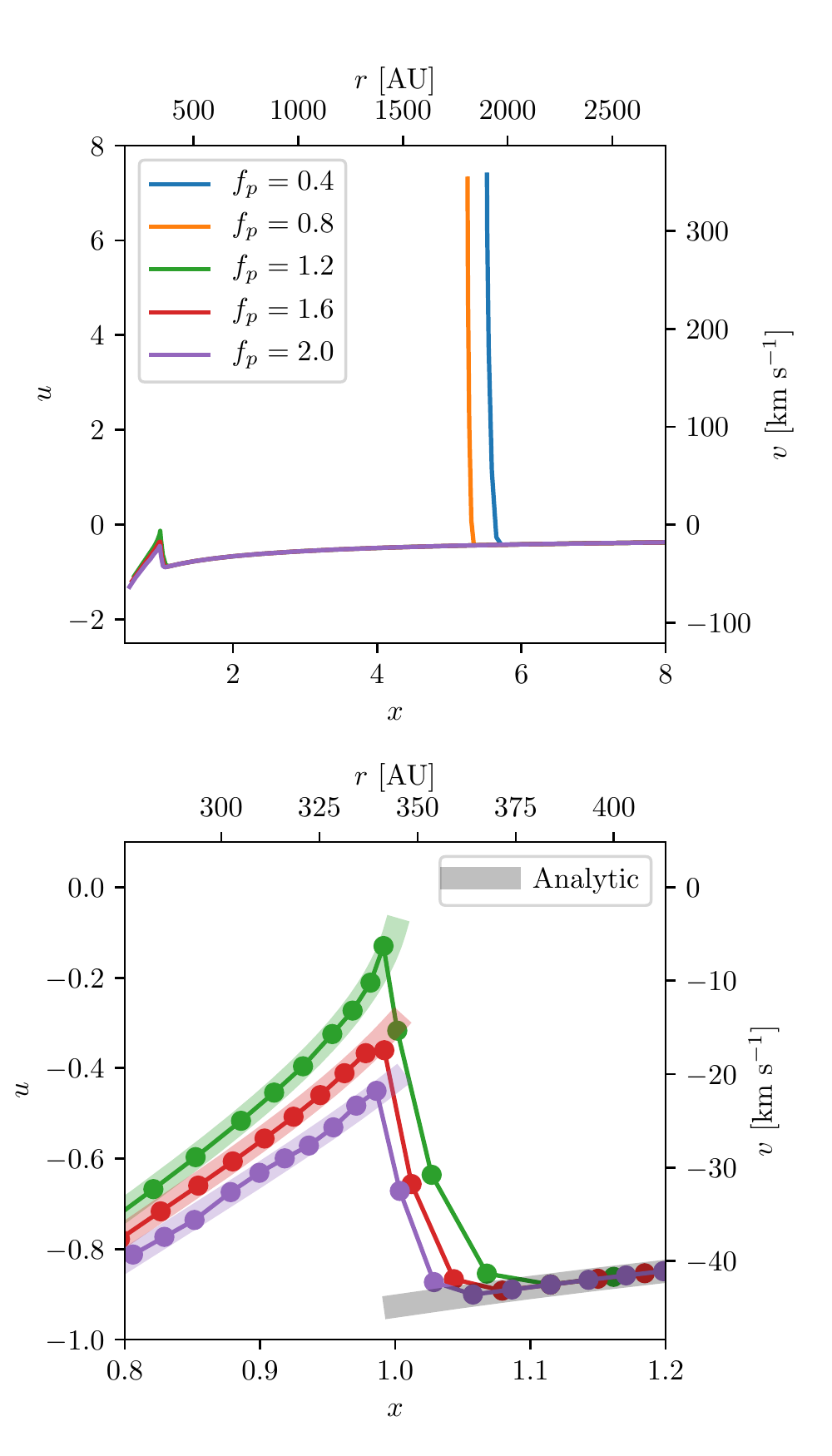}
\caption{
\label{fig:highres}
Results of simulations of radiation-inhibited accretion flows with $\Delta m = 10^{-6}$, sufficient to resolve the flow inside the dust destruction front by $\approx 25$ cells. Both panels show dimensionless velocity $u$ versus dimensionless radius $x$ at a dimensionless time $s=1$; the top and right axes show the corresponding dimensional velocity and radius for a central object luminosity $L = 10^6$ $L_\odot$. Different colours indicate different values of $f_p$ (or equivalently mass accretion rate), as indicated in the legend. The top panel shows a large part of the simulation domain, while the bottom panel zooms in around the shock as the dust destruction front. In the bottom panel, circles represent individual zone edges, and thick lines in the background show the analytic solution for the indicated value of $f_p$. 
}
\end{figure}

The analytic expectation for these simulations is that the cases $f_p = 1.2$, $1.6$, and $2.0$ should result in steady inflow, since all of these are above $f_{p,\rm crit}$, while in the cases $f_p = 0.4$ and 0.8 the inflow should be reversed by radiation pressure. \autoref{fig:highres} shows the results of the simulations, and demonstrates that they successfully recover this analytic result. For the cases that should produce continuous accretion flows, the velocity as a function of position shows near-perfect agreement with the analytic solution, with the sole exception that the shock at $x=1$ has been broadened to $\approx 4$ zones in width by the artificial viscosity. In the two cases where accretion flow should be reversed, the simulation has indeed produced a shock moving outward at high velocity, which prevents accretion and sweeps up the accretion flow into a dense shell.

\subsubsection{Varying resolution}

Having verified that the code easily reproduces the analytic solution at high resolution, I now investigate its ability to do so at lower resolution. I repeat the $f_p = 1.2$ run (using $f_p = 1.6$ or $2$ produces qualitatively identical results) using mass resolutions of $\Delta m = 10^{-6}$ (identical to the case shown in \autoref{fig:highres}), $10^{-5}$, $10^{-4}$, $10^{-3}$, and $10^{-2}$. The mass interior to the dust sublimation front for $f_p=1.2$ is $m_s = 1.9\times 10^{-5}$, so at the highest resolution the simulation resolves this region by $\approx 19$ zones, and at the lowest resolution the mass of a single zone is $\approx 500$ times the mass inside the dust sublimation radius. I initialise the simulations exactly as in \autoref{sssec:high_res}, using 20,000 cell edges for the run with $\Delta m = 10^{-6}$, 2,000 for the run with $\Delta m = 10^{-5}$, and 1,000 cell edges for all other cases; the different numbers of resolution elements are to ensure that all mass does not accrete onto the central sink before the end of the simulation, even when the mass resolution is very high. I run all simulations to $s=1000$. For a central source luminosity $L = 10^6$ $L_\odot$ and the light to mass ratio of a zero age stellar population ($\Psi = 1.1\times 10^3$ $L_\odot/M_\odot$), the run time is 33.7 kyr, and the mass resolution range is $9.1\times 10^{-4}$ $M_\odot$ to $9.1$ $M_\odot$

\begin{figure*}
\includegraphics[width=\textwidth]{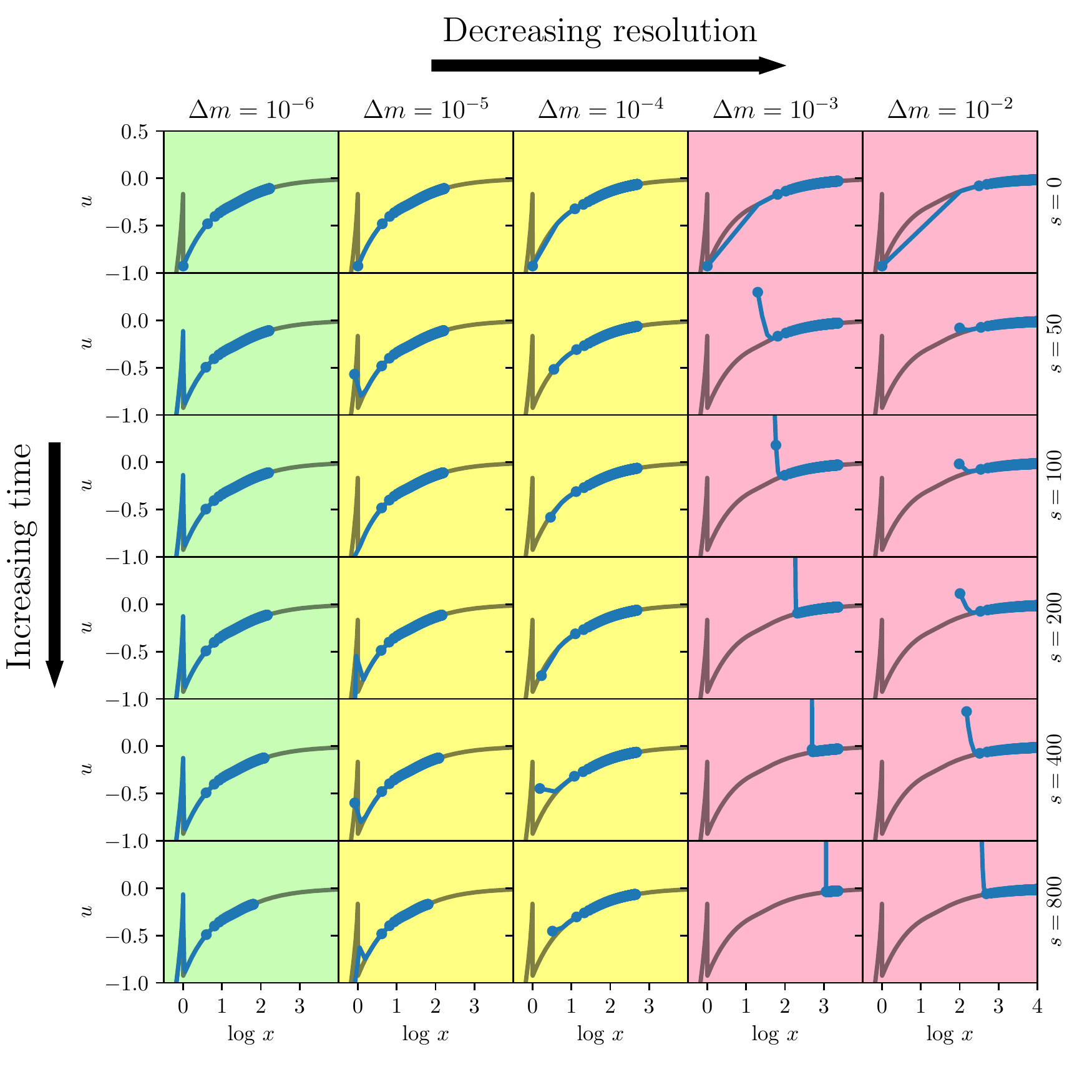}
\caption{
\label{fig:res_study}
Results of simulations of radiation-inhibited accretion flows with varying resolution. Each panel shows a plot of dimensionless cell edge position $x$ versus dimensionless velocity $u$. Grey lines show the analytic solution, while blue lines with circles show the simulation result. Circles indicate individual zone edges, with 200 edges plotted rather than all edges in order to minimise clutter, but the corresponding lines show all zones. The different columns show simulations with mass resolutions from $\Delta m = 10^{-6}$ to $\Delta m = 10^{-2}$, as indicated at the top of the column. Background colours qualitatively indicate how well a given simulation resolves the dust sublimation zone mass $m_s = 1.9\times 10^{-5}$: the green colour indicates good resolution, $m_s / \Delta m > 10$, yellow indicates marginal resolution, $0.1 < m_s/\Delta_m < 10$, and red indicates poor resolution, $m_s / \Delta m < 0.1$. Different rows show the results at different times, from $s=0$ (top, initial condition) to $s = 800$, as indicated by the labels on the right of each row.
}
\end{figure*}

\autoref{fig:res_study} shows the results of the resolution study. In the figure, resolution decreases to the right, from $\Delta m = 10^{-6}$ in the left column to $\Delta m = 10^{-2}$ in the right column. Time increases downward, with $s=0$ in the top row and $s=800$ in the bottom row. For resolutions of $\Delta m = 10^{-6}$, $10^{-5}$, and $10^{-4}$, the qualitative result is correct: a steady-state inflow.\footnote{However, note that the $\Delta m = 10^{-4}$ case is marginal. For the test runs shown in \autoref{fig:res_study} it is qualitatively similar to the $\Delta m = 10^{-5}$ case, but small changes in numerical procedure (e.g., a different CFL number or a different artificial viscosity coefficient) can cause it to behave like the $\Delta m = 10^{-3}$ case instead.} The effects of decreasing resolution are essentially as one might naively expect: the shock at the dust destruction front is well-resolved at $\Delta m = 10^{-6}$, but for $\Delta m = 10^{-5}$ and $10^{-4}$, its effects are only marginally visible as a slight upturn in the velocity of the innermost zones that comes and goes in time. However, in the region that is resolved the inflow matches the analytic solution extremely well.

For $\Delta m = 10^{-3}$ and $10^{-2}$, on the other hand, the results are completely, qualitatively different. In those cases the radiation pressure is able to drive a shock outward that produces a thin shell, much as occurred in the high resolution simulations with $f_p < f_{p,\rm crit}$. By $s=800$, the shock has swept up all the material interior to $x\approx 1000$, shutting off accretion and creating an evacuated zone that is $\approx 10^9$ times the volume of the true dust destruction front. The dependence on resolution is obvious: the simulations that resolve the dust destruction front, at least marginally (recall $\Delta m = 10^{-4}$ corresponds to the mass resolution being $\approx 5$ times the mass inside the sublimation front), produce qualitatively correct solutions, while simulations that do not resolve the dust destruction front produce qualitatively incorrect results.

Moreover, there is no evidence for convergence in the simulations that fail to resolve the dust destruction front. The swept-up shell has advanced to a larger radius and at a higher speed in the higher resolution run than in the lower resolution one. The important conclusion to draw from this is that, when the dust destruction front is unresolved, the results need not converge smoothly toward the true solution. Instead, they may actually move \textit{away} from the true solution as resolution improves, until the dust destruction front is finally resolved and the true solution is recovered.

\subsection{Analysis: why do low resolution simulations produce incorrect results?}
\label{ssec:analysis}

To understand why the low resolution simulations fail, it is helpful to examine the process of momentum deposition in the accretion flow. For any given (dimensionless) force per unit mass $f$, producing an acceleration $du/ds = f$, the total amount of momentum per unit mass that the force delivers to a Lagrangian fluid element as it falls from infinity to $x$ is
\begin{equation}
\Delta p = -\int_x^\infty \frac{du}{ds} \frac{ds}{dx} \, dx = \int_x^{\infty} \frac{f}{|u|} \, dx.
\end{equation}
We can use this expression, together with the analytic inflow solution derived in \autoref{ssec:inflow_sol}, to evaluate the total momentum deposited by gravity, infrared radiation, and direct stellar radiation. We can then compare this to the momentum deposition for fluid elements calculated directly from the simulations.

\begin{figure}
\includegraphics[width=\columnwidth]{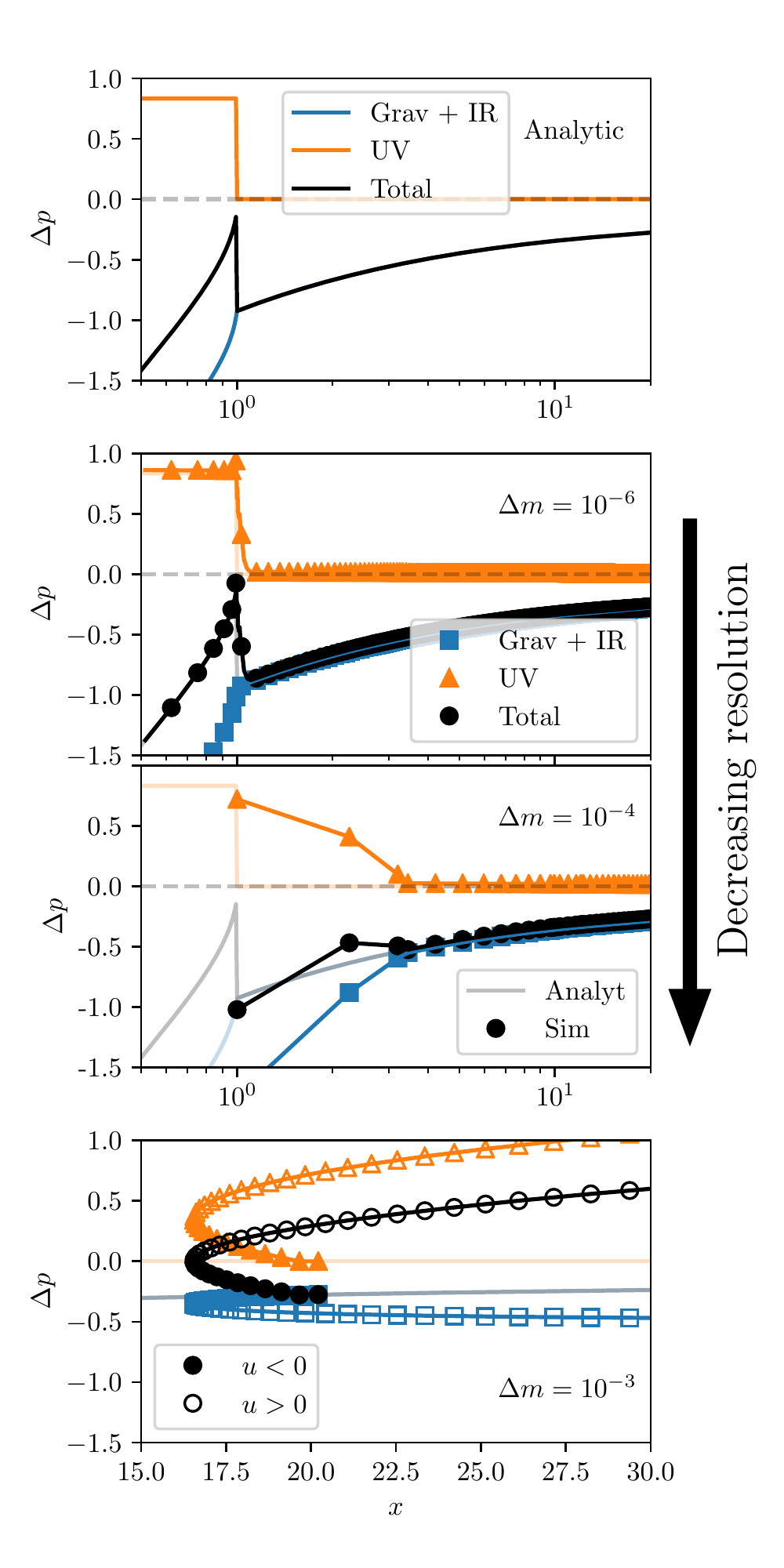}
\caption{
\label{fig:p_vs_x}
Cumulative momentum $\Delta p$ deposited by various forces during the time it takes a fluid element falling from infinity to reach position $x$. The top panel shows the analytic result, with lines indicating momentum deposition by gravity (blue), stellar UV radiation (orange), and the sum of the two (black). The faint dashed line shows $\Delta p = 0$ for reference. I reproduce this analytic solution in the background of the bottom three panels for comparison. The middle two panels show numerical results for a sample cell edge drawn from the simulations at resolutions of $\Delta m = 10^{-6}$ and $10^{-4}$, as indicated, both of which produce the qualitatively correct result. For the simulations, the plotted stellar UV momentum deposited also includes the contribution from pressure and viscous forces, which mediate the shock. The bottom panel shows the numerical result for a simulation with a resolution $\Delta m = 10^{-3}$, which yields the qualitatively incorrect result that inflow reverses; note the difference in $x$ axis range, since the flow in this case never reaches $x=1$. In this plot, filled symbols indicate times when the gas is inflowing, while open symbols indicate times when the flow has reversed and the gas is outflowing.
}
\end{figure}

I plot the analytically-computed momentum deposition versus radial position $x$ in the upper panel of \autoref{fig:p_vs_x}. The figure shows that the inward gravitational force, reduced by the outward infrared radiation force, delivers momentum to a fluid parcel steadily as it moves inward. The majority of the momentum is delivered at radii close to the central source. This is opposed by the stellar radiation field, which delivers no momentum at all until the fluid element reaches $x\approx 1$, where it provides a sharp impulse $1/f_p$. For the chosen simulation parameters, which have $f_p$ slightly above the critical value, the impulse is slightly smaller than the total momentum deposited by gravity up to $x \approx 1$. Thus the total momentum deposition remains negative at all radii, and inflow occurs.

The middle panels of \autoref{fig:p_vs_x} show the momentum deposition computed numerically for a single Lagrangian point drawn from the $\Delta m = 10^{-6}$ and $\Delta m = 10^{-4}$ simulations, both of which recover the qualitatively correct result that inflow occurs. In the highest resolution case the momentum deposition as a function of position agrees extremely well with the analytic solution. For $\Delta m = 10^{-4}$, where the dust sublimation front is not quite resolved, we see that the location where the UV force delivers its momentum is $x\approx 2$ rather than $x = 1$. This is simply a reflection of the fact that, since the dust sublimation zone is not quite resolved, the first cell edge that has non-zero opacity is located somewhat beyond $x=1$ rather than almost at $x=1$. The physical consequence of this displacement is that, at the point where the gas first encounters the direct stellar radiation force, its inward momentum is slightly less than in the $\Delta m = 10^{-6}$ case. This difference is not enough to allow the radiation pressure to reverse the inflow, but it clearly moves the system in that direction.

The bottom panel of \autoref{fig:p_vs_x} shows the numerical result for a sample point from the simulation with a resolution $\Delta m = 10^{-3}$. In the initial setup for this simulation the first cell edge is at $x=1$, and this point flows into the central object as it should. However, the outer edge of this first cell is at $x\approx 20$, where the momentum it carries is smaller by a factor of $u(1)/u(20) \approx 3.4$ than what it would have upon reaching the dust sublimation front at $x=1$. However, since this is the first zone outside the dust sublimation front (once the innermost zone moves inside $x=1$ and thus becomes transparent), this is where the stellar radiation field deposits its its momentum. While the stellar radiation field does not carry enough momentum to overcome the momentum of inflow at the dust sublimation front, it \textit{is} sufficient to overcome the factor of $\sim 3$ lower momentum that the flow carries at this larger radius. Consequently, the flow reverses. From that point on, the problem only gets worse. The reversed flow begins to move away from the central radiation source, which does not change the amount of momentum the direct stellar radiation deposits, but \textit{does} further reduce the amount of momentum that gravity is able to deliver to fluid elements before they confront the stellar radiation field. This leads to a runaway outward-moving shell.

We can therefore summarise the nature of the problem in low-resolution simulations as follows: gravity and radiation forces deposit momentum with very different spatial distributions. Gravity delivers momentum smoothly, with most of the impulse occurring on the smallest scales, while radiation (at least direct stellar radiation) delivers a fixed amount of momentum per unit time as a sharp impulse wherever it is absorbed. This asymmetry means that these two forces behave very differently when they are effectively softened by low resolution. Stellar radiation delivers the same amount of momentum; it simply deposits that momentum in a location that is further from the source. On the other hand, gravity delivers less momentum overall, because the small scales on which it should deposit momentum are unresolved. In a radiation-inhibited accretion flow, where the qualitative outcome is determined by the balance between gravitational and radiation forces, the result of making an error in the \textit{location} of radiative force but in the \textit{quantity} of gravitational force is that low-resolution simulations can be disastrously incorrect, and greatly overestimate the effectiveness of radiation.

This analysis also suggests a final point. In the 1D spherical numerical scheme I have used, I have been careful to implement the discretised radiation force so as to guarantee that the rate of radially-outward momentum deposition is correct to machine precision, regardless of the resolution. However, this is only possible in a spherically symmetric geometry. In a more general geometry, in a calculation that does not resolve the dust destruction front where the momentum should be deposited, there is no way to design a discrete scheme that guarantees that exactly the right amount of radial momentum will be deposited.  \citet{hopkins18a} point out the existence of a pathological case where the total radial momentum deposited cancels to zero exactly because all the momentum deposition happens inside a single cell, but this is only an extreme version of a more generic problem, and neither their proposed fix nor any possible alternatives to it can guarantee that the radial momentum deposited per unit time will be exactly $L/c$. Depending on the sign and magnitude of the error in the radial momentum, it is possible that any particular numerical scheme might make an error in the amount of radial radiation momentum deposited that is even larger than the error it makes due to softening of gravity. If this happens, then the scheme will tend to underestimate rather than overestimate the effectiveness of radiation. Thus in a general 3D geometry, errors in both directions are possible.

\section{A subgrid model for radiation feedback}
\label{sec:subgrid_model}

How can one fix this problem? The best solution is simply to resolve the dust sublimation front, as is routinely done in simulations of the formation of individual stars or star clusters. Simulations that are able to reach spatial resolutions of $\sim 100$ AU or better (for Eulerian methods), or mass resolutions of $\sim 0.01$ $M_\odot$ or better (for Lagrangian ones), need no modification.

However, resolving the dust sublimation front is impractical in simulations on the scales of entire giant molecular clouds, galaxies, or cosmology. Recall that, for the dimensional scalings that correspond to a luminosity of $10^6$ $L_\odot$ and the mass to light ratio of a zero age stellar population, the lowest-resolution test presented in \autoref{sec:simulations} that yielded a qualitatively-correct result had a mass resolution of $\approx 0.09$ $M_\odot$, and that case was marginal; to have confidence in the result, one would likely want a mass resolution closer to $0.01$ $M_\odot$. This is significantly better than the highest resolution simulations of individual molecular clouds presented by \citet{grudic18a} or \citet{kim18a}, and orders of magnitude beyond state of the art galactic or cosmological simulations \citep[e.g.,][]{hopkins11a, agertz13a, hopkins18b, kannan18a}.

If one cannot resolve the dust destruction front, then a next-best approach is to implement a subgrid model that accounts for the balance of radiation and gravity on unresolved scales, using the analytic results summarised in \autoref{ssec:analytic_summary}. Recall that the key parameter that determines whether radiation pressure is able to reverse inflow, or whether all the radiation momentum will be advected back into the radiation source, is the momentum flux of the accretion flow. In a simulation that does not resolve the dust sublimation front one cannot calculate this directly, because the momentum is mostly added to the flow on unresolved scales. However, the \textit{mass flux} of the accretion flow does not change from large to small scales. Thus one can measure the resolved mass flux and then use this to estimate the unresolved momentum flux simply by assuming that the gas accelerates under gravity (reduced by infrared radiation pressure) toward the dust sublimation front. If the estimated momentum flux at the dust sublimation front is above the critical value (i.e., $f_p > f_{p,\rm crit}$), then the direct radiation pressure force should be turned off, on the grounds that any momentum carried by it will be advected back into the star and will not escape to scales that are resolved in the simulation. If the momentum flux is below the critical value, the radiation force can be applied on resolved scales normally, under the assumption that radiation will be able to reverse the accretion flow and escape. To be precise, in dimensional terms the model is that the direct radiation momentum per unit time deposited in the simulation by a source of luminosity $L$ should be
\begin{equation}
\label{eq:acc_quench}
\frac{dp}{dt} = 
\left\{
\begin{array}{ll}
0, & \dot{M} > 6.5\times 10^{-4} L_6^{3/4}\,M_\odot\,\mathrm{yr}^{-1} \\
L/c, & \dot{M} \leq 6.5\times 10^{-4} L_6^{3/4}\,M_\odot\,\mathrm{yr}^{-1} 
\end{array}
\right.,
\end{equation}
where $L$ is the luminosity of the source, $L_6 = L/10^6$ $L_\odot$, $\dot{M}$ is the mass accretion rate onto the source, and the numerical coefficients are based on Milky Way-like dust. Dependence on metallicity or other dust properties can be added by plugging the relevant dust parameters into \autoref{eq:fp_con}. If the sharp transition from zero to non-zero force is numerically problematic, it can be replaced with a suitably-smoothed function of $\dot{M}$ instead.

The method by which the mass flux $\dot{M}$ is estimated will depend on the nature of the simulation. In simulations using sink particles \citep{bate95a, krumholz04a, federrath10a, gong13a}, the accretion rate is built into the sink particle method, and one can take it directly from that. In simulations that do not use sink particles (most galaxy-scale and cosmological simulations), a reasonable estimate is
\begin{equation}
\label{eq:mdot_est}
\dot{M} = \max\left[-4\pi \left(\frac{\Delta x}{2}\right)^2 \nabla\cdot(\rho \mathbf{v}), 
4\pi \frac{G^2 M_*^2}{c_s^{3/2}} \rho, \sqrt{G\rho^3}\Delta x^3 \right],
\end{equation}
where $\rho$, $\mathbf{v}$, and $c_s$ are the density, velocity, and sound speed in the simulation evaluated for the resolution element in which the radiation source is located, and $\Delta x$ is the simulation resolution. The first term represents the estimated mass flux onto the resolution element containing the radiation source (and may be computed in a variety of ways depending on the hydrodynamic scheme), the second represents the accretion rate produced by the gravity of the point mass (the \citet{bondi52a} rate), and the third is the rate of mass accretion that should be produced by the self-gravity of the gas within the resolution element.

To test the utility of this subgrid model, I implement it in the simple simulation code described in \autoref{ssec:scheme}. Since the code is spherical Lagrangian, there is no resolution element that includes the point mass, and thus only the first of the three conditions in \autoref{eq:mdot_est} applies. I compute the estimated mass accretion rate from the density, velocity, and position of the first Lagrangian cell:
\begin{equation}
\dot{m} = -4\pi \left(\frac{x_0+x_1}{2}\right)^2 \left(\frac{u_0+u_1}{2}\right) b_{i+1/2}.
\end{equation}
In the dimensionless units used in the simulations, the condition for the accretion flow to shut off the UV radiation feedback is
\begin{equation}
\frac{f_\tau}{4\pi f_E\eta} \dot{m} > f_{p,\rm crit}.
\end{equation}
I test the model by running a series of simulations initialised as in \autoref{sssec:high_res} using momentum fluxes $f_p = 0.25$, $0.5$, $1$, $2$, and $4$. The analytic solution is that accretion should be reversed for $f_p = 0.25$, $0.5$, and $1$, and should continue despite radiation pressure for $f_p = 2$ and 4. I test whether simulations with mass resolutions $\Delta m = 10^{-3}$, $10^{-2}$, $10^{-1}$, and $10^{0}$ can recover this result; for a central object with luminosity $10^6$ $L_\odot$ and the light to mass ratio of a zero age stellar population, these dimensionless mass resolutions correspond to physical mass resolutions $0.9$, $9$, $90$, and $900$ $M_\odot$, respectively. Thus $\Delta m = 10^{-3}$ corresponds to a resolution that could plausibly be achieved in a simulation of a single giant molecular cloud, while $\Delta m = 10^{0}$ corresponds to a resolution that might be achievable for a simulation of an isolated galaxy or a cosmological zoom-in. I use 1000 Lagrangian points for each simulation, and I run for a time $s=10^{2.5}$, $10^{3.5}$, $10^{4.5}$, and $10^{5.5}$ at the four different resolutions; for the fiducial dimensional scaling, the run times are from $11$ kyr to $0.11$ Myr.

\begin{figure*}
\includegraphics[width=\textwidth]{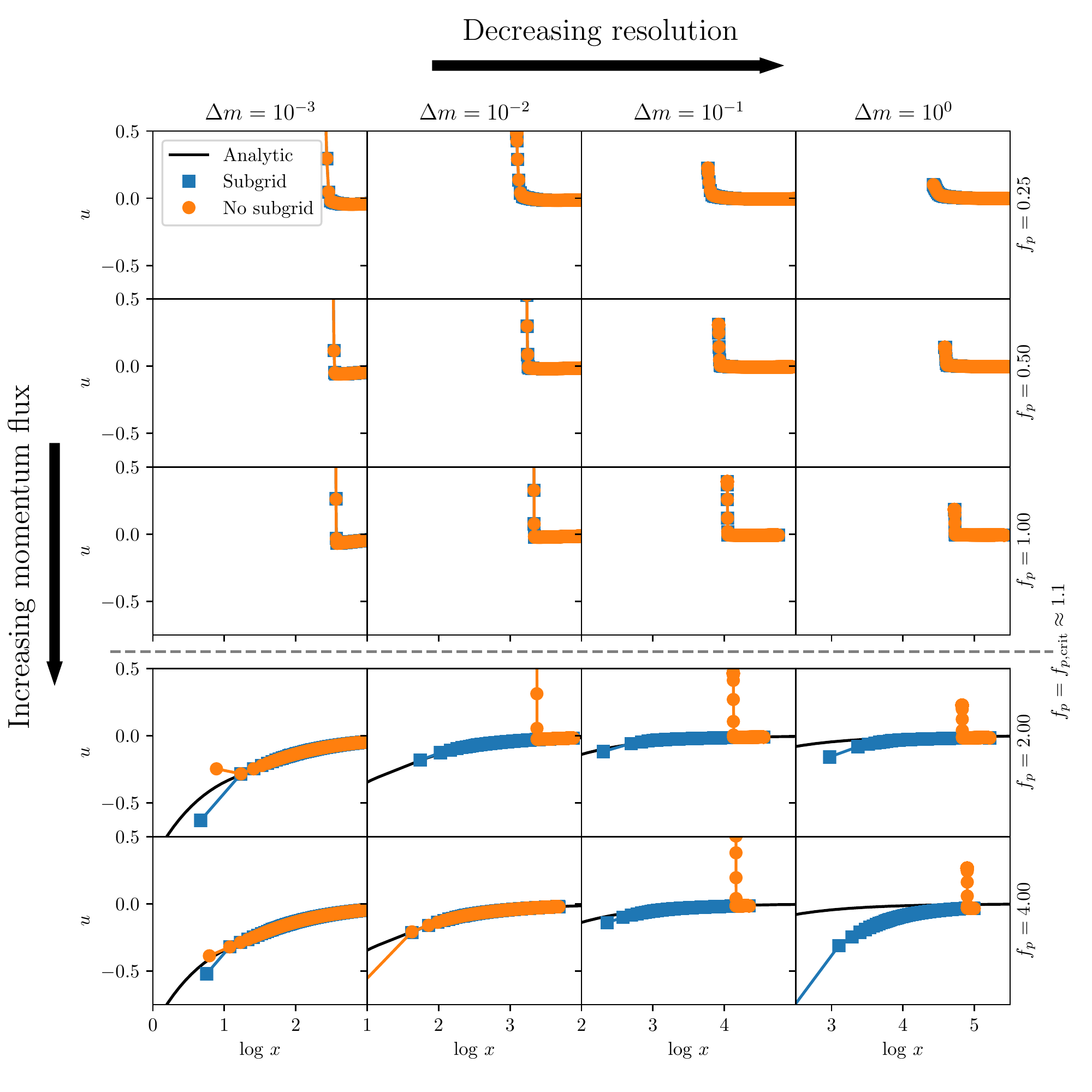}
\caption{
\label{fig:subgrid}
Tests of the subgrid model. Each panel shows velocity $u$ versus position $x$ for Lagrangian points at the end of two simulations of a radiatively-inhibited accretion flow: one using the subgrid model (blue squares) and one run without it (orange circles). The columns, from left to right, show simulations with mass resolutions of $\Delta m = 10^{-3}$ to $10^0$, as indicated at the tops of the columns. The rows, from top to bottom, show simulations where the accretion flow carries different momentum fluxes, from $f_p = 0.25$ to $4.0$ as indicated to the right of the rows. The dashed grey horizontal line separates the simulations with $f_p < f_{p,\rm crit}$, for which the correct answer is that radiation reverses the accretion flow and drives a shell outward, from those with $f_p > f_{p,\rm crit}$ for which the correct answer is that accretion is not halted. In the rows with $f_p > f_{p,\rm crit}$, the black line shows the analytic solution for the accretion flow. Note that the simulation using the subgrid model deviates shows faster inflow than the analytic solution at the lowest mass resolution; this is because the low resolution simulation reaches size scales large enough that gas self-gravity is no longer negligible, contrary to the assumption of the analytic solution. The gas self-gravity leads to faster infall.
}
\end{figure*}

\autoref{fig:subgrid} shows the state of each simulation at the final time. For the reasons discussed in \autoref{ssec:analysis}, the simulations without subgrid models often show that radiation pressure is able to reverse inflow even when it should not be able to, with the problem getting worse as the resolution decreases. At the resolutions typical of galaxy-scale simulations, as shown in the right two columns, radiation pressure is able to choke off accretion even though it carries a factor of $\approx 4$ too little momentum to do so in reality. As noted above, standard convergence tests, where one runs the same initial conditions at a range of resolutions, are unlikely to detect this problem. Consider the non-subgrid case for the row $f_p = 2$: the results are nearly identical at mass resolutions of $\Delta m = 10^{-2}$, $10^{-1}$, and $10^0$.\footnote{Astute readers may notice that the shock position is not the same in each panel, but recall that, due to the differing resolutions, the simulations in the three panels have run for different amounts of time. If one examines the shock position as a function of time, it is qualitatively similar in all three runs.} It is only once the dust sublimation front begins to be resolved that the solution switches over to the correct one. The system does in fact converge, but the nature of the convergence is that the value of $f_p$ at which the solution switches from inward accretion to outgoing shock converges to the correct value as the resolution increases. A test of a single set of initial conditions, with a single $f_p$ value, will not detect this effect unless the resolutions chosen happen to straddle the resolution at which, for that $f_p$, the resolution at which the character of the solution changes.

The runs using the subgrid model show no such problems. When the momentum flux is below the critical value, $f_p < f_{p,\rm crit}$, the runs with the subgrid model are completely identical to those without it, as they should be. For momentum fluxes above the critical value, the simulations using the subgrid model recover the correct solution at all resolutions. In summary, the subgrid model appears to resolve the problem of overestimation of the effectiveness of radiation feedback in low resolution simulations.

That said, it is important to notice that there is an additional a practical difficulty in using this model for cosmological  simulations, as opposed to higher-resolution simulations of isolated galaxies or individual molecular clouds. As noted in \autoref{ssec:analytic_summary}, in situations where there are multiple radiation sources separated by $\gg r_s$, the condition for accretion to quench radiative feedback (\autoref{eq:acc_quench}) should be calculated individually for each point source. In simulations where the typical ``star particle" has a mass of $\sim 100$ $M_\odot$ or less, typical of isolated galaxy or molecular cloud simulations, this is not a problem, because a stellar population of this size is likely to have its light output dominated by the single most massive star \citep[e.g.,][]{da-silva12a}. Thus it is reasonable to treat each ``star particle" in the simulation as a single point source and apply \autoref{eq:acc_quench} to it. However, if the resolution is such that individual ``star particles" have masses $\gtrsim 10^3$ $M_\odot$ and thus represent clusters of stars with many individual luminous sources that contribute non-negligibly to the total luminosity, as is usually the case for cosmological simulations, then one cannot simply plug the total accretion rate and the total luminosity into \autoref{eq:acc_quench}. Instead, one will require a further subgrid model for the luminosities of the individual sources that comprise the star particle, and for how the accretion rate $\dot{M}$ is likely to be partitioned between them.

\section{Summary}
\label{sec:conclusion}

\subsection{Implications}

In this paper I show that the structure of dusty accretion flows impeded by radiation pressure imposes an important resolution limit on numerical simulations. A simulation that does not resolve the dust sublimation front, which for stellar radiation sources lies at distances of $\sim 100 - 1000$ AU, will generally overestimate the effectiveness of radiation forces in halting the accretion flow. The physical origin of this resolution criterion is easy to understand. Whether radiation forces are able to halt an accretion flow comes down to a contest of momenta: radiation reverses the accretion flow if it delivers more outward momentum to the gas than gravity provides inward momentum. If gravity wins the contest, the radiation delivers all its momentum to gas that is then advected back onto the central star, so accretion continues and radiation feedback has no effect on the flow at larger distances. 

However, gravity and radiation deposit their momenta in very different spatial patterns. Thanks to the very high opacity of dusty gas to radiation at the colour temperature of a star, the radiation field delivers a sharp impulse with a fixed momentum flux wherever it is absorbed. Gravity, on the other hand, delivers momentum slowly as gas falls, with most of the momentum delivered on the smallest scales. As a result of this difference, the outcome of a contest between these two forces depends critically on where it takes place, with gravity more likely to win when the gas is able to fall farther toward the radiation source before encountering its direct radiation field. In real accretion flows the contest between gravity and radiation occurs at the dust sublimation front, the location closest to a star where the dust opacity is high enough to absorb all the stellar momentum in a thin layer. In a simulation that does not resolve the dust sublimation front, however, the location of the contest is artificially moved outward to the smallest resolved scales, thereby making it too easy for radiation to reverse the accretion flow and for radiation momentum to escape to large distances. The problem grows worse as the resolution does. However, conventional convergence tests will not easily reveal this, because the solution switches sharply from outflow to inflow when the momentum flux of (resolution-softened) gravity goes from smaller to larger than that of radiation. Unless the convergence test happens to catch the resolution where this switch happens, the results may appear converged even when they are not.

This numerical problem is very likely the origin of the surprising divergence in results between simulations of molecular clouds or galaxies, where at least some authors report that radiation momentum feedback is very effective at halting accretion and preventing star formation, and simulations of the formation of individual massive stars systems, which invariably show that radiation does not lead to low star formation efficiency despite the fact that single massive stars have higher light to mass ratios than IMF-averaged stellar populations. The single star simulations resolve the dust destruction front, while the molecular cloud and galaxy ones do not. This explanation of the discrepancy suggest that least some published work should be re-examined. While it is probably not feasible to resolve the dust sublimation front in molecular cloud or galaxy simulations, a second-best option is to use the subgrid model I develop in \autoref{sec:subgrid_model} to account for momentum deposition by gravity on unresolved scales, and shut off the radiation momentum deposition in the regime where advection should prevent it from reaching scales that are resolved in the simulation.

\subsection{Future prospects}

While my analysis in this paper has mostly focused on feedback from direct stellar radiation, the numerical issue I have identified is likely relevant for other types of momentum-limited feedback as well. All of these have in common the property that the outcome depends sensitively on the amount of momentum carried by the accretion flow where it encounters the feedback, and thus they are vulnerable to error when low resolution artificially moves the encounter radius outward. An obvious example is infrared radiation pressure; while this force does not deliver its momentum on quite as small a scale as direct radiation pressure, it is nonetheless a force where the momentum deposition is largest in small regions where higher radiation temperature produces higher opacity. As pointed out by \citet{crocker18a}, the common numerical practice of using the high infrared opacity that applies in these small regions on the much larger scales probed by low-resolution simulations is likely to artificially favour feedback over gravity in much the same way I have explored here.

The flip side of this point is that any feedback mechanism that is able to create some ``standoff distance" between stellar radiation sources and accretion flows is likely to be much more effective than one might initially suspect. For example, if a population of newly-formed stars were to launch stellar winds that created a small bubble of hot gas $\sim 0.1$ pc ($\approx 2\times 10^4$ AU) in radius around themselves, that by itself would probably not be very significant for a galaxy- or molecular cloud-scale simulation, which would struggle to resolve such small scales. Physically, however, this could in principle make it much easier for stellar radiation feedback to reverse the accretion flow, by moving the radius at which the accretion flow hits the radiation field outward by a factor of $\approx 10-100$ compared to a case without a stellar wind. It is therefore urgent that we carry out simulations to study these effects by simultaneously resolving the dust sublimation front and including more than one type of feedback.

\section*{Acknowledgements}

I thank R.~M.~Crocker, M.~Y.~Grudi\'c, C.~F.~McKee, E.~C.~Ostriker, S.~Rei{\ss}l, M.~A.~Skinner, and T.~A.~Thompson for helpful discussions and comments on the manuscript, and the referee, J.~Rosdahl, for an insightful referee report. Support for this work was provided by the Australian Research Council (Discovery Projects award DP160100695, and the Centre of Excellence for All Sky Astrophysics in 3 Dimensions, project CE170100013), and by the National Computational Infrastructure (NCI), which is supported by the Australian Government.

\bibliographystyle{mn2e}
\bibliography{refs}

\end{document}